\documentclass[prb,floatfix,showpacs,12pt]{revtex4}
\usepackage{graphicx}

\begin{document}

\title{ Effect of topology on the transport properties of two interacting dots }

\author{ V.M. Apel$^1$, Maria A. Davidovich$^1$, E.V. Anda$^1$, G. Chiappe$^2$ and C.A.Busser$^3$}
\affiliation{$^1$Departamento de F\'{\i}sica, Pontif\'{\i}cia
Universidade Cat\'olica do Rio de Janeiro, 22453-900,  Brazil \\
\it $^2$Departamento de F\'{\i}sica,  Facultad de Ciencias Exactas
y Naturales, Universidad de Buenos Aires, 1428, Argentina \\
\it $^3$ National High Magnetic Field Lab, Florida State University, Tallahassee, FL 32306, USA }

\begin{abstract}
The transport properties of a system of two interacting dots, one of them  directly connected to the leads constituting a side-coupled configuration (SCD), are studied in the weak and strong tunnel-coupling limits. The conductance behavior of the SCD structure has new and richer physics than the better studied system of two dots aligned with the leads (ACD). In the weak coupling regime and in the case of one electron per dot, the ACD 
configuration gives rise to two 
mostly independent Kondo states. In the SCD topology, the inserted dot is in a Kondo state while  
the side-connected one presents Coulomb blockade properties. Moreover, the dot spins change their behavior, from an antiferromagnetic coupling to a ferromagnetic correlation, as a consequence of the interaction with the conduction electrons. 
The system is governed by the Kondo effect related to the dot that is embedded into the leads. The role of the 
side-connected dot is to introduce, when at 
resonance, a new path for the electrons to go through giving rise to the  interferences responsible for the suppression 
of the conductance. These results depend on the values of the intra-dot Coulomb interactions. In the case where the many-body interaction is restricted to the side-connected dot, its Kondo correlation is responsible for the  scattering of  the conduction electrons giving rise to the conductance suppression.

\end{abstract}
\pacs{73.63.-b,73.63.Kv}
\maketitle 

I. INTRODUCTION

In the last years interest in double-quantum-dot systems has been renewed since they have been 
proposed as basic building blocks for quantum computers\cite{DiVicenzo}. The strength of the 
interaction between the two quantum dots determines the character of the electronic states and
the nature of the transport through them. In the limit of weak tunneling interaction the 
electrons are localized on the individual dots while in the strong limit the delocalized 
electronic charge is no longer quantized. The transport properties depend also on the topology 
of the double-dot system. For the case of the two dots aligned with the leads (ACD), in the weak coupling limit the
interaction of each individual dot with the conduction electrons of the nearby lead gives rise 
to the Kondo
phenomenon. In the opposite limit, the strong inter-dot tunneling interaction together with the
intra-dot Coulomb
repulsion result in an anti-parallel correlation of the two dot spins that quenches the Kondo 
effect. As the tunneling
interaction is increased the dot system goes continuously from the Kondo regime to an 
antiferromagnetic state.
The interplay between the interactions that give rise either to the Kondo effect or to the 
dot-dot antiferromagnetic correlation is reflected on the conductance of the system 
\cite{Georges e nos1}.
Another system of two-interacting dots in the side-connected configuration (SCD) has special interest since it 
permits, in principle, the control of the current along the leads by changing the state of 
charge of the side-connected dot. The conductance of this system, in the particular 
case where the Coulomb interaction is restricted to the side-connected dot, has already 
been investigated in two cases: a)considering the charge of the side-connected dot fixed at one 
electron and neglecting charge fluctuations so that the dot could be treated as an $S=1/2$ 
spin\cite{Hersh}; b)letting its charge to be varied by the application of a gate potential
\cite{nos2}. These studies conclude that at low temperature the linear-response conductance in 
the leads is suppressed due to the Kondo scattering of spin $S=1/2$ of the side-connected dot.
However, new and very interesting physics appears when both dots have many-body interactions and both charges can be varied. Manipulating the dot gate potentials, the system can go through a ground state transition modifying its properties, particularly the conductance and the interdot and the dot-conduction-electron spin correlations.  

In this 
paper we address this new physics by studying the conductance for all states of charge of the dots and various regimes of the system, in the weak and strong 
inter-dot tunneling coupling. We analyze the differences between the SCD and the  ACD configuration. We find that the transport 
properties for these two topologies are quite different when the inter-dot tunneling coupling is weak. In the SCD configuration, if there is one electron at each dot, their spins are ferromagnetic correlated constituting a triplet state, while the dot inserted into the leads forms a conduction electron Kondo cloud in its neighborhood and the side-connected one stays in the Coulomb blockade regime.

II. METHOD

An Anderson two-impurity first-neighbor tight-binding Hamiltonian represents the system schematically depicted in Fig. 1. The total Hamiltonian reads,

\begin{eqnarray}
H &=& \sum_{ r=\alpha,\beta\atop\sigma}\left( V_{r} +
\frac{U_{r}}{2} n_{r \bar\sigma}\right)n_{r\sigma}+ 
t_{D}\sum_{\sigma} (c^+_{\alpha\sigma} c_{\beta\sigma}\! + \!{\rm c.c.})\nonumber\\
&+& t'\sum_{\sigma} \left[ (c^+_{\alpha\sigma} c_{1\sigma}\! +
\! c^+_{\alpha\sigma} c_{\bar1\sigma}\! +\!{\rm c.c.})\right] +  t \sum_{i,j,\sigma}c^+_{i\sigma}c_{j \sigma} \label{H} \end{eqnarray}

where dot $\alpha$, inserted into the leads, interacts with its nearest neighbor  sites $1$ and $\bar1$ through 
the hopping matrix element $t'$. The interactions between the dots $\alpha$ and $\beta$ and among the lead sites are represented by the hopping matrix elements $t_D$ and $t$, respectively.
The parameters $V_r$ and $U_r$ represent, respectively, the gate potential and the intra-dot 
Coulomb interaction in each dot.
To describe the very low temperature properties we calculate the one particle Green
functions $G_{\alpha,\beta}$ at the dots. Part 
of the system, consisting of a cluster that includes the two dots and some lead sites, is 
exactly solved and then embedded into the rest of the contacts. The Green functions are made to 
satisfy a Dyson equation $\hat G=\hat g+%
\hat g \hat T \hat G$ where $\hat g$ is the cluster Green function matrix
and $\hat T$ is the matrix of the coupling Hamiltonian between the cluster
and the rest of the system. The undressed 
Green function $\hat g$ is
calculated using the cluster ground state obtained by the Lanczos method\cite{Lanc}. In
order to guarantee consistency the charge of the dressed and undressed
cluster is imposed to be the same. We calculate $\hat g$ as a combination of
the Green function of $n$ and $n+1$ electrons with weight $1-p$ and $p$, $\hat %
g = (1-p)\hat g_n + p \hat g_{n+1}$. The charge of the undressed cluster is $%
q_c=(1-p)n + p(n+1)$. The charge of the cluster when linked to
the leads can be expressed as $Q_c=2\int^{\epsilon_F}_{-\infty}\sum_i
Im\,Gii(\omega)d\omega$, where $i$ runs over all the cluster sites. This
equation plus the condition $q_c=Q_c$ constitute a system
of two equations which requires a self-consistent solution to obtain $p$ and
$n$\cite{Valeria}. Using the Keldysh\cite{Keldysh} formalism the conductance can be
written as
\begin{equation}
G={\frac{e^2t^2 }{h}}|G_{\alpha,\alpha}|^2[\rho(\epsilon_F)]^2
\end{equation}
where $G_{\alpha,\alpha}$ is the Green function at dot $\alpha$
while $\rho(\epsilon_F)$is the density of states at the Fermi level at the
first neighbors of dot $\alpha$, when disconnected from it.

This approximation has shown to be very accurate, almost numerically exact when the cluster is 
of the size of the Kondo cloud $hv_F/T_K$, where $v_F$ is the Fermi velocity, although it
gives qualitatively reliable results even for shorter clusters. Moreover, the procedure 
satisfies the Luttinger-Ward
identity that ensures the fulfillment of the Friedel sum-rule and of the Fermi liquid 
properties.

We calculate the conductance of the system in the weak and strong 
inter-dot coupling limits, as the gate potentials applied to the
dots are changed. The density of states, the charge inside the dots and the various spin-spin correlation functions are also
calculated in order to characterize the state of the system.

III.   RESULTS

All energies are in units of the bandwidth $W$ and the Fermi level is chosen to be 
$\epsilon_F = 0$. The other parameter values are $U_{\alpha}= U_{\beta}=0.5$, and 
$\Gamma = t'{^2}/W =0.015$. Let us first analyze the weak tunneling coupling case, where
we take $t_D = 0.15$. A general view of the conductance is
presented in Fig. 2 for the whole range of energy of the
states localized at the dots, $V_\alpha$ and $V_\beta$. 
 The conductance presents
quite different characteristics depending upon the localization in the two gate potentials parameter space. In order to study these phenomena we obtain the conductance modifying the gate potentials in three different ways. When  $V_\alpha$ is changed while
maintaining constant $V_\beta$ (along an horizontal line in the figure),
the conductance presents only one peak. If, on the contrary,
$V_\alpha$ is maintained at a fixed  value in the range $-0.5 <
V_\alpha < 0$, and $V_\beta$ is varied (along a vertical line in the
figure), the conductance is almost constant except around two
values of $V_\beta$ where it cancels out. This reflects the
asymmetry of the two dots in the topology we are studying. On the
other hand, if both dot energies are simultaneously varied, for
example, along the diagonal $V_\alpha = V_\beta$ as shown in
the figure, the conductance possesses three peaks. In what follows we discuss
these cases in more detail.

The conductance, dot charges and spin correlations for the case where the side-connected dot
local energy $V_\beta$ is fixed, are presented in Fig. 3 as a function of the energy 
$V_\alpha$. We consider two situations: $V_\beta = -0.25$, shown in Fig. 3, when dot $\beta$ has
just one electron and $V_\beta = -1.0$ when it has two. 
In both cases the conductance and the dot $\alpha$ charge and spin correlation with the conduction electrons are similar. The conductance shows one peak with width of the order of $U_\alpha$ in the 
region $-0.5 < V_\alpha < 0$, where the charge at dot $\alpha$ grows continuously from zero to 
two, and the spin-spin correlation between dot $\alpha$ and conduction electrons, 
$< S_{\alpha}, S_c >$, is negative. These results characterize a Kondo state related to dot 
$\alpha$. The  total spin of the two dots, $S_T$, and the spin correlation between dot $\beta$ and 
conduction electrons, $< S_{\beta}, S_c >$, differ for the two situations as expected, since 
dot $\beta$ with two electrons has no spin, what is consistent with $S_T = 0.35$ and 
$< S_{\beta}, S_c > =0$. For the case where each dot has just one electron we obtain 
$S_T = 0.65$ and $< S_{\beta}, S_c > < 0$, indicating a ferromagnetic dot-dot  correlation and an anti-parallel alignment of dot $\beta$ and the conduction electron spins.
 In order to characterize the many-body state of the dots we represent in 
the inset of Fig. 3 the density of states (DOS) projected on each dot, at the electron-hole 
symmetry condition $V_\alpha = V_\beta = -0.25$. The DOS 
at dot $\alpha$ presents a resonance that is pinned at the Fermi level as  $V_\alpha$ is 
changed, confirming that this dot has a Kondo resonance. However, the DOS at dot $\beta$ is 
zero at the Fermi level and has two peaks separated by $U_\beta$, indicating that dot $\beta$ 
is in the Coulomb blockade regime.

It is well known that the spins of two sites with large intra-Coulomb repulsion $U$ interacting 
through an inter-site matrix element $t_D$ are coupled by an antiferromagnetic interaction of 
the order 
of $t_D^2/U$. If the two dots we are analyzing were to have a dominant antiferromagnetic spin 
interaction the system would not have a net spin to couple  
Kondo-like with the conduction electrons. We conclude that in the weak coupling limit the system reduces 
its energy by coupling the spin of the embedded dot in a Kondo-like manner and by transforming  
the inter-dot antiferromagnetic spin-spin correlation into a renormalized 
ferromagnetic interaction. For this to occur the energy gained by the Kondo 
ground state, of the order of the Kondo temperature, $T_K$, has to be greater than the energy 
lost due to the ferromagnetic coupling between the 
dots. $T_K$ has to satisfy the relation $T_K > t_D^2/U$. This can be clarified by solving a 
three site problem, representing the 
conduction electrons and the interacting dots, in a topological disposition such that only 
one of the dots interacts with the 
conducting site through $t'$.
The dot-conduction electron interaction and the dot-dot interaction $t_D$ break in a different manner
the degeneracy of the S=1
and S=0 spin states of the two dots. These states represent the ferromagnetic and antiferromagnetic situation, 
respectively. 
While the 
the energy of the S=1 state is renormalized by a value of the order of the Kondo 
temperature of a localized two site system\cite{Fulde}, given by $2 t'^2/V_\alpha$, 
the renormalization energy of the S=0 state is of the order of $t_D^2/U $ corresponding 
to the inter-dot antiferromagnetic coupling. The interplay of these two quantities determines the ground state of the system. 
As a conclusion, in the regime we are analyzing the spins of dots $\alpha$ and $\beta$ are 
parallel-aligned giving rise to a total spin greater than $1/2$.

Since dot $\alpha$ forms a Kondo state its spin is anti-parallel 
to the conduction electron spins, and so does the spin of dot $\beta$. However, dot $\beta$ is not in the Kondo regime.   
In spite of having a localized spin, when out of resonance the side-connected dot plays no role in 
the conductance of the 
system that is governed uniquely by the Kondo effect of the inserted dot.

The results obtained when the role played by the gate potentials is interchanged, $V_\beta$ is 
varied while maintaining $V_\alpha = -0.25$, are shown in Fig 4.
Since dot $\alpha$ has a constant number of electrons equal to one, it is in the 
Kondo regime for any value of $V_\beta$. However the conductance is not constant but is 
suppressed at two values of $V_\beta$. This happens around $V_\beta =0$ and 
$V_\beta = U_\beta$, 
when the side-connected dot levels cross the Fermi energy, adding a new path 
for the electrons to go through, resulting in a Fano interference. As one electron enters 
into dot $\beta$, at $V_\beta = 0$, the total spin of the dots increases from $S_T = 0.35$ to 
$S_T = 0.64$ and maintains this value up to the entrance of the second electron at 
$V_\beta = U_\beta = 0.5$.  In this same region of gate potential the spin correlation 
$< S_{\beta}, S_c >$ results to be negative, in agreement with the interpretation given 
before that the two dot spins are parallel aligned. Except for the discontinuities, the 
charge at dot 
$\beta$ is maintained about constant over the Coulomb blockade region, as expected.

The case where both dot energies are simultaneously changed is depicted in Fig. 5, for 
$V_\alpha = V_\beta$. The conductance shows three peaks and cancels out at two values of the 
gate potential due to the same interference phenomenon as discussed just before. 
Although 
dot $\alpha$ is in the Kondo regime the conductance has zeros due to destructive interference 
between two electron paths.
Charge enters continuously into dot $\alpha$, as expected for a Kondo state, and abruptly at dot 
$\beta$, where it remains constant and equal to one for the whole range of gate 
potential, reflecting the Coulomb blockade regime of the dot.
The behavior of the total spin and the spin-spin correlations as a function of gate potential is similar to the one presented in Fig. 3. 

The discontinuities present in the charge of dot $\beta$, the total spin and the spin 
correlation, seen in Figs. 4 and 5 for two values of the gate potential, are due to a ground state crossing that results in an abrupt change
of the properties of the system. These two ground states have different spatial parity reflecting the inverse
symmetry that the system possesses. The discontinuity in the physical magnitudes acquires a 
cross-over behavior when this symmetry is broken. This can be shown by simply modifying one of the coupling matrix elements of the embedded dot.

      We consider now the strong coupling limit by taking the inter-dot tunneling interaction 
$t_D = 0.9$. The conductance, dot charges and spin correlation as a function of the 
gate potentials, for the case $V_\alpha = V_\beta$, are depicted in Fig. 6. In this limit, 
$t_D > t'$, the two dots have an effective anti-ferromagnetic coupling that  aligns 
their spins anti-parallel as discussed above, destroying the Kondo effect. The results are 
similar to the ones obtained for the ACD configuration.
 The similarity comes from the fact that, in this limit, the two dots act as an entity so that 
the way each dot is connected to the leads is not relevant.
 
IV. SUMMARY

In conclusion, in the weak inter-dot tunneling regime the transport properties of a system of two 
interacting dots connected to leads have quite different features according to its topology. 
In the SCD configuration studied here, when the Kondo temperature is greater than the antiferromagnetic interdot spin interaction, the spins are ferromagnetic
correlated due to the interaction of one of them with the conducting electrons. In this case the dot embedded into the 
leads forms a Kondo state with the conduction 
electrons while the side-connected dot is in the Coulomb blockade regime. The conductance is 
governed by the Kondo effect and shows three peaks as both gate 
potentials are varied. The 
suppression is due to a Fano interference of the electrons, whenever the Coulomb blockade
peaks of the side-connected dot cross the Fermi level. On the other hand, in the ACD 
configuration the dots are independently Kondo correlated to the conduction electrons of the leads to 
which they are connected\cite{Georges e nos1}. The conductance shows a broad peak due to the Kondo 
resonance at both dots. 

In the strong-coupling limit the behavior of the conductance is the same for the 
two topologies. The similarity comes from the fact that in this limit the two dots act as a 
molecule so that the way each dot is connected to the leads is not relevant.  

Therefore, if both dots have many-body interactions the side-connected dot does not give any 
Kondo contribution to the transport properties of this system. 
This result is quite different from the one obtained in the case the many-body interaction is 
restricted to the side-connected dot, where it is just the Kondo effect of this dot that 
scatters the conduction electrons giving rise to the conductance cancellation\cite{Hersh}.

We acknowledge FAPERJ, CNPq, CAPES and Buenos Aires University (grant UBACYT) for financial support.

\newpage

\begin{figure}
\includegraphics[width=8cm]{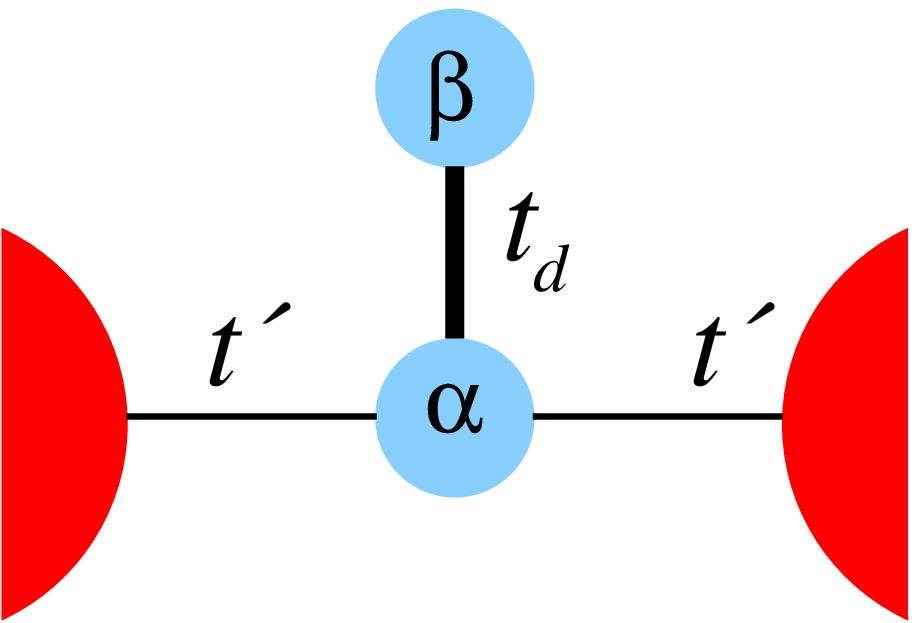}
\caption{Two interacting dots in the side-connected configuration.} 
\end{figure}

\begin{figure}
\includegraphics[width=8cm]{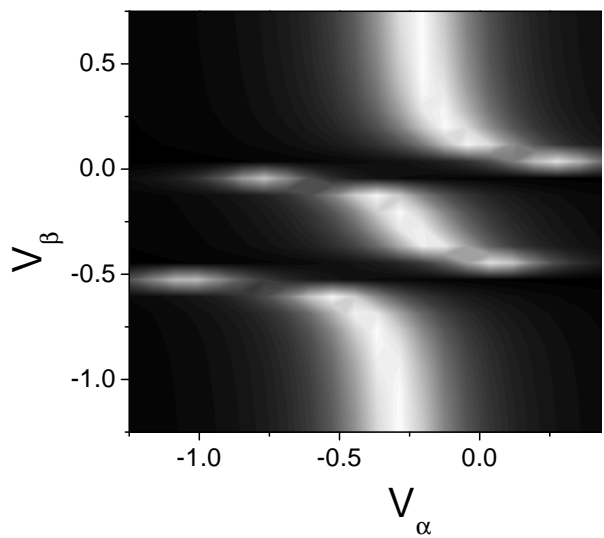}
\caption{Conductance (white, maximum; black, minimum) as a
function of the gate potential $V_\alpha$ at the aligned dot and
$V_\beta$ at the side-connected dot. Weak coupling limit.}
\end{figure}

\begin{figure}
\includegraphics[width=8cm]{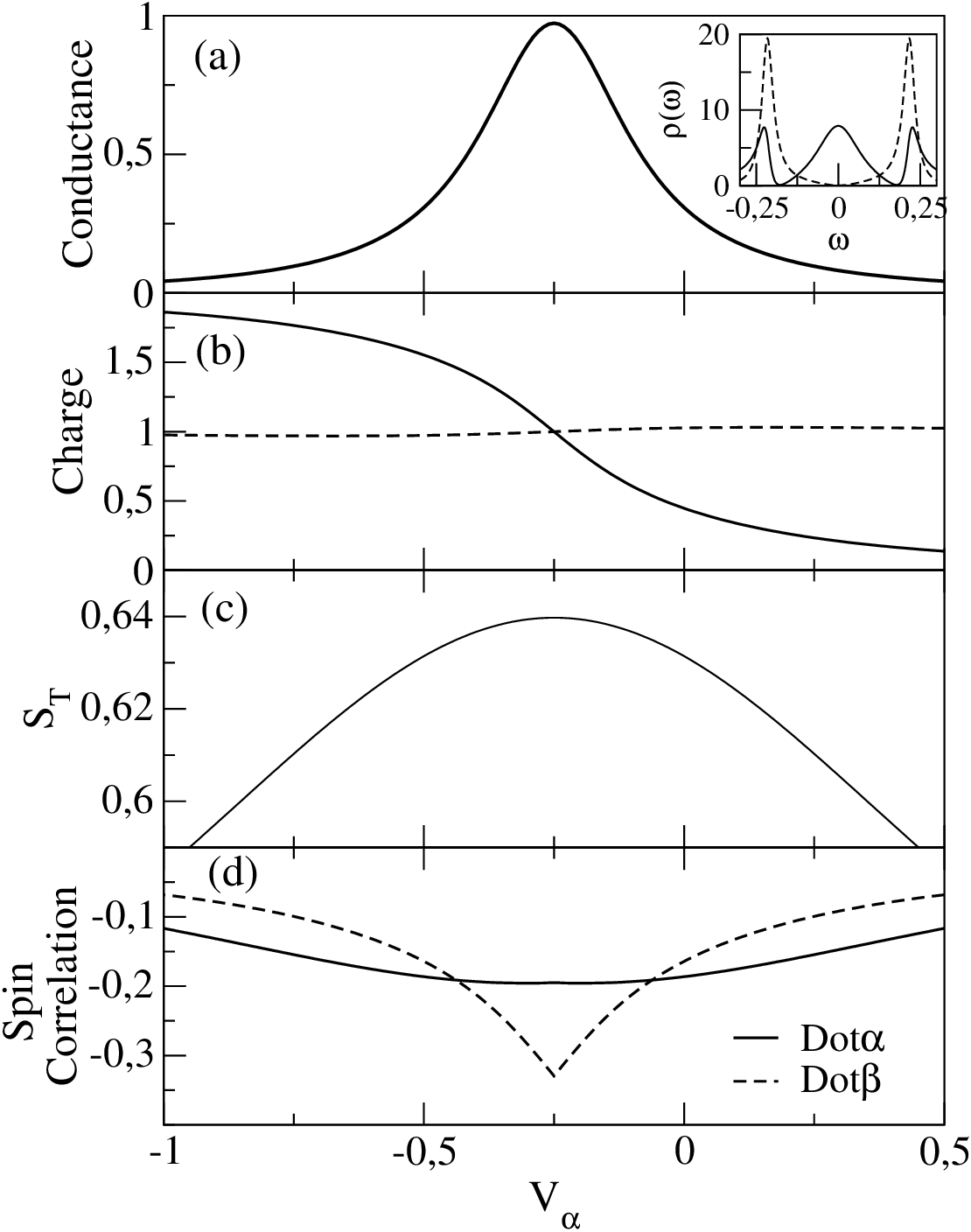}
\caption{Weak-coupling results for $V_{\beta} = -0.25$, as a function of $V_{\alpha}$. (a)Conductance (in units of $e^{2}/h$), (b)charge (in units of {\it e})at dot $\alpha$ (continuous line) and dot $\beta$ (dashed line), (c)total spin and (d)spin correlation $< S_{\alpha}, S_c >$ (continuous line) and $< S_{\beta}, S_c >$ (dashed line). }
\end{figure}

\begin{figure}
\includegraphics[width=8cm]{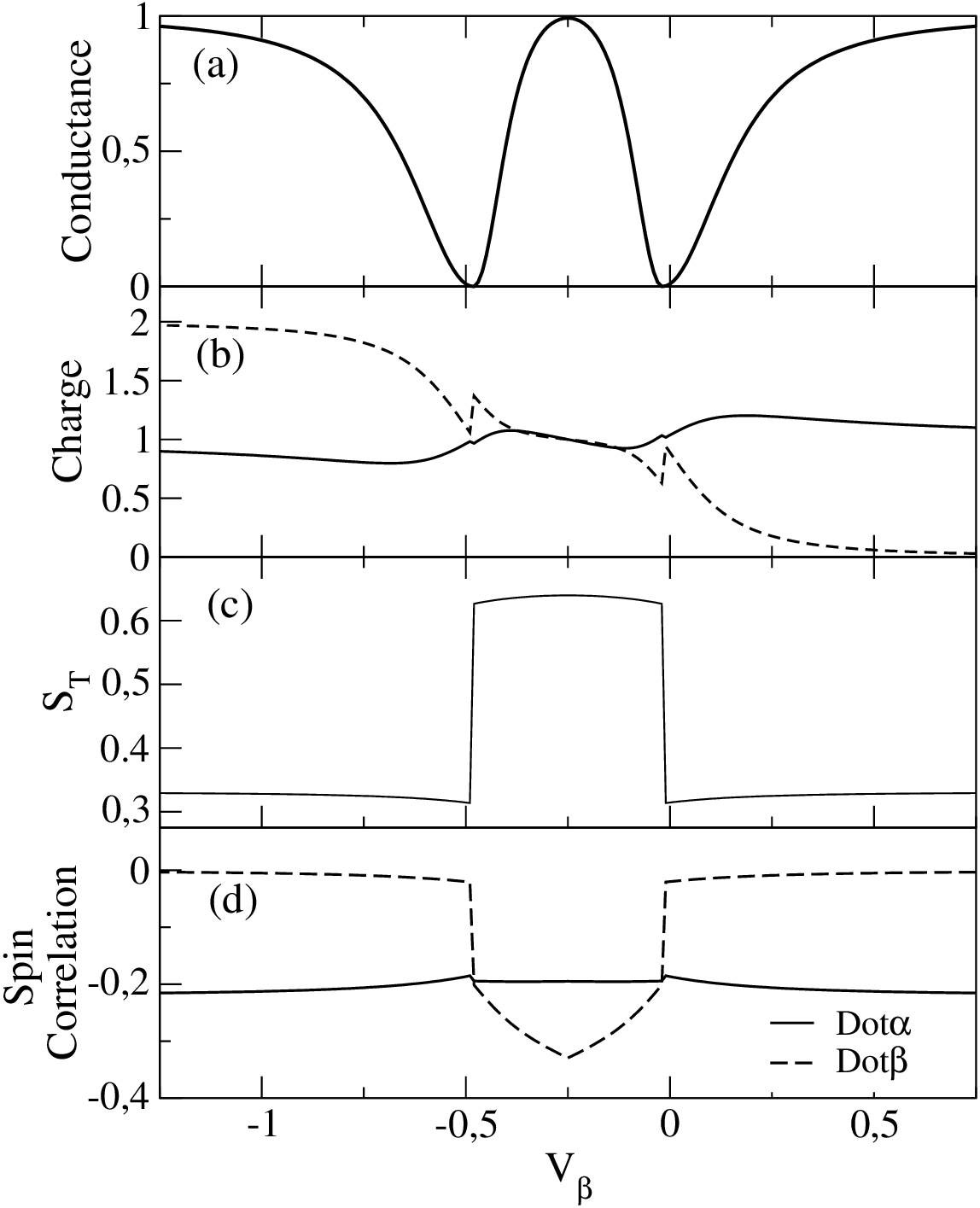}
\caption{ Weak-coupling results for $V_\alpha = -0.25$, as a function of $V_\beta$. (a)Conductance (in units of $e^{2}/h$), (b)charge (in units of {\it e})at dot $\alpha$ (continuous line) and dot $\beta$ (dashed line), (c)total spin and (d)spin correlation $< S_{\alpha}, S_c >$ (continuous line) and $< S_{\beta}, S_c >$ (dashed line).}
\end{figure}

\begin{figure}
\includegraphics[width=8cm]{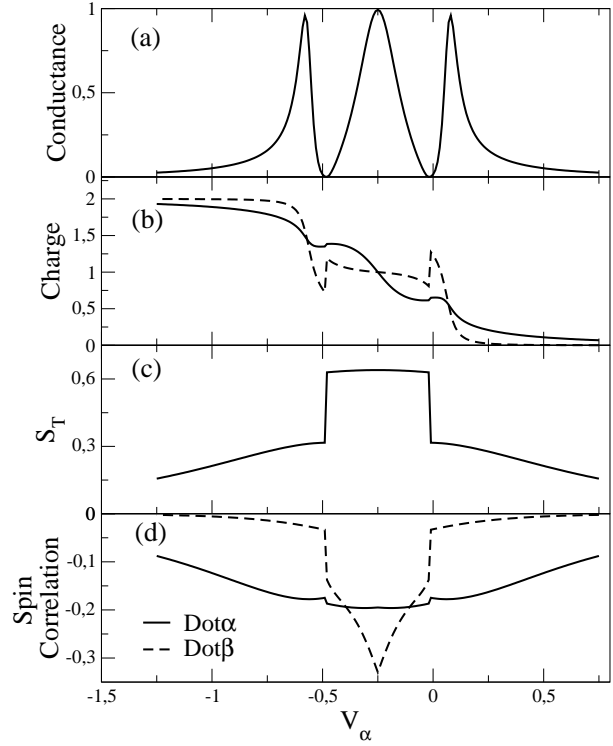}
\caption{ Weak-coupling results as a function of gate potential for $V_\alpha = V_\beta$. (a)Conductance (in units of $e^{2}/h$), (b)charge (in units of {\it e})at dot $\alpha$ (continuous line) and dot $\beta$ (dashed line), (c)total spin and (d)spin correlation $< S_{\alpha}, S_c >$ (continuous line) and $< S_{\beta}, S_c >$ (dashed line).}
\end{figure}

\begin{figure}
\includegraphics[width=8cm]{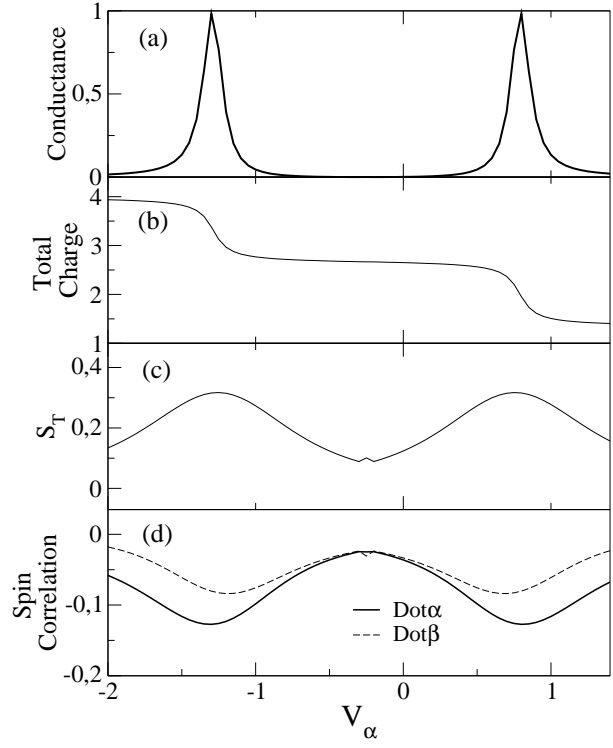}
\caption{ Strong-coupling results as a function of gate potential for $V_\alpha = V_\beta$. (a)Conductance (in units of $e^{2}/h$), (b)total charge (in units of {\it e}), (c)total spin and (d)spin correlation $< S_{\alpha}, S_c >$ (continuous line) and $< S_{\beta}, S_c >$ (dashed line).}

\end{figure}

\end{document}